\documentclass[epj,final]{svjour}

\usepackage{latexsym}
\usepackage{url}
\usepackage{amsfonts}
\usepackage{amsmath, amssymb}
\RequirePackage{graphicx, subcaption}

\begin{document}
\title{Cosmic voids and induced hyperbolicity. II. Sensitivity to void/wall scales}
\author{M.~Samsonyan\inst{1}, A.A.~Kocharyan\inst{2}, A.~Stepanian\inst{1}, V.G.~Gurzadyan\inst{1,3}
}                     
%
%
\institute{Center for Cosmology and Astrophysics, Alikhanian National Laboratory and Yerevan State University, Yerevan, Armenia  \and 
School of Physics and Astronomy, Monash University, Clayton, Australia \and
SIA, Sapienza Universita di Roma, Rome, Italy}
\date{Received: date / Revised version: date}
%

\abstract{Cosmic voids as typical under-density regions in the large scale Universe are known for their hyperbolic properties as an ability to deviate the photon beams. The under-density then is acting as the negative curvature in the hyperbolic spaces.  The hyperbolicity of voids has to lead to distortion in the statistical analysis at galactic surveys. We reveal the sensitivity of the hyperbolicity and hence of the distortion with respect to the ratio of void/wall scales which are observable parameters. This provides a principal possibility to use the distortion in the galactic surveys in revealing the line-of sight number of cosmic voids and their characteristic scales.} 
%
\maketitle

\section{Introduction}

The cosmic voids as the under-dense regions in the large scale matter distribution attract much attention due to their principal role in revealing of the early evolution of the Universe, see \cite{P1,Pan,Ham,Con,Din} and references therein. The physical characteristics of the voids, e.g. the distribution of void/wall scales, shapes, under-density parameter, the void galaxy vs wall galaxy correlations, etc, are informative tests for the theory of density perturbation evolution, approaches to the nature of dark matter and dark energy, modified gravity models, etc., e.g.\cite{Hoy,Ce,Nad,Fl,Cap}. 

The concept of a local void has been associated with the Local Supercluster structure \cite{GC,LR,NCK} and to the dynamics of the galaxies in the vicinity of the Local Group \cite{RG,GS21,Cai}. Another remarkable structure, the Cold Spot - the non-Gaussian region in Cosmic Microwave Background (CMB) sky map - has been associated to a large cosmic void \cite{spot,Sz,Hig2,Rag,Vi}. Within the context of our study below, it is important to note that, the void nature of the Cold Spot was concluded in \cite{spot} analyzing the {\it Planck's} microwave data using the Kolmogorov stochasticity parameter as a quantitative measure of the property of hyperbolicity of photon beams caused by the voids as divergent lenses \cite{GK1,GK2}, see also \cite{Das}. 

The hyperbolicity of geodesic flows in negatively curved spaces is studied in the theory of dynamical systems \cite{An,Arn}. In the context of voids it was shown \cite{GK1,GK2} that the under-density of voids leads to hyperbolicity of photon beams even for flat and positively curves spaces. Then, in \cite{SK} the anisotropy of photon beams was shown also to contribute in inducing hyperbolicity of the geodesic flows.

Continuing the study of the property of photon beam hyperbolicity and particularly the image distortion caused by that effect, we now reveal its sensitivity on the ratio of the void vs wall scales. Since the dimensions of the voids and of their walls are observable quantities, this opens a principal possibility to detect the image distortion due to the effect of hyperbolicity for individual voids in dependence on those parameters. Also, it will allow, having the distortion detected in certain galactic surveys, to reveal the number of line-of-sight voids and their mean void/wall scales. Particularly, the void hyperbolicity/distortion can contribute to the distortion of galaxy correlation functions already detected at galactic surveys \cite{Pea,Guz,Mc}.       

\section{Hyperbolicity of geodesic flows}

The hyperbolicity of geodesic flows in d-dim Riemannian manifold $M$ is defined via the Jacobi equation of the deviation of close geodesics \cite{An,Arn}
\begin{equation}
\frac {d^2\parallel n \parallel ^2}{ds^2}=
     -2K_{u,n}\parallel n \parallel ^2
    + 2\parallel \nabla_{u} n \parallel ^2,
\end{equation}
where $K_{u,n}$ is the two-dimensional curvature 
\begin{equation}
K_{u,n}=\frac{<Riem(n,u)u,n>}
{\parallel u \parallel ^2\parallel n \parallel ^2 - <u,n>^2} =\\
\frac{<Riem(n,u)u,n>}
{\parallel n \parallel ^2}.
\end{equation}
The Riemannian curvature $Riem(n,u)$ is 
\begin{eqnarray*}
& &
Riem(n,u)u=([\nabla_{n},\nabla_{u}] - \nabla_{[n,u]}) u =
[\nabla_{n},\nabla_{u}] u,\\
& &
\nabla_{n} u - \nabla_{u} n = [n,u] = 0,
\end{eqnarray*}
and the deviation vector {\bf n} is orthogonal to the velocity vector {\bf u}
\begin{equation}
<n,u>=0.
\end{equation}
 
The exponential deviation of geodesics as the signature of hyperbolicity
\begin{equation}
\parallel n(s)\parallel \geq \frac{1}{2}
\parallel n(0)\parallel \exp (\sqrt {-2k} s), \qquad s>0,
\end{equation}
occurs at the negative two-dimensional curvature
\begin{equation}
k=\max_{u,n} \{K_{u,n}\}<0.
\end{equation}
			
We will inquire into the dependence of the hyperbolicity of photon beams caused by the cosmic voids on the physical parameters of the voids and walls, i.e. on the under-density and over-density parameters, respectively, and their spatial scales. We reveal the sensitivity of the distortion induced by the hyperbolicity to the ratio of the void/wall scales. 

\section{Hyperbolicity vs distortion}
 
The averaged Jacobi equation for (d+1)-dimensional Lorentzian manifold can be written as \cite{GK1}
\begin{equation}\label{gdev}
\frac{d^2\ell}{d \eta^2}+ \frac{\mathfrak{r}}{d-1}\,\ell =0\,,
\end{equation}
where $\mathfrak{r}$ is the d-dimensional spatial Ricci scalar. 

The line element for Friedmann-Lemaitre-Robertson-Walker  metric with small perturbation $\phi$ is ($c=1$)
\begin{equation}\label{FLRW}
ds^2 = -(1+2\phi) dt^2 + (1- 2\phi) a^2(t) d\gamma^2\,, 
\end{equation}
where $k$ is the sectional curvature of spatial geometry, and $d\gamma^2$ denotes the spherical ($k=1$), Euclidean ($k=0$) or hyperbolic ($k=-1$) geometries. 

The following conditions are assumed to be fulfilled for the perturbation $\phi$  \cite{Holz}
\begin{equation}\label{Pert}
|\phi|\ll1, \quad \left(\frac{\partial \phi}{\partial t}\right)^2\ll a^{-2} ||\nabla\phi||^2\, .
\end{equation}

At weak-field limit of Eq.(\ref{FLRW}) the Ricci scalar has the form ($d=3$)
\begin{equation}\label{RicciL}
\frac{\mathfrak{r}}{2}
= k + 2 H_0^2\Omega_{m}\ \Theta\ \tilde{\delta},
\end{equation}
where the following definition is used
$$
\tilde{\delta}=\frac{\delta\rho}{a\langle\rho\rangle}
=\frac{\rho-\langle\rho\rangle}{a\langle\rho\rangle}
$$ 
and 
\begin{equation}\label{theta}
\Theta= \frac{3}{4}\left(1+ \frac{(\nabla^2\phi,\mathbf{u}\otimes\mathbf{u})}{\Delta\phi}\right)\,.
\end{equation}
Here the function $\Theta$ reflects the anisotropy of the photon beams, so that $\Theta=1$ corresponding to spherical distribution, $\langle\mathbf{u}\otimes\mathbf{u}\rangle=\frac{1}{3}\mathbf{\gamma}$.

\noindent 
Our aim is the study of sample matter configurations distributed periodically, that is voids and walls \cite{GK2}. Then, Eq.(\ref{gdev}) will take the form \cite{SK}
\begin{equation}
\frac{d^2\ell}{d \eta^2}+ \frac{\mathfrak{r}}{2}\ell =0\,,
\end{equation}

For $k=0$ one has
\begin{equation}
\frac{d^2\ell}{d \eta^2}+ 2 H_0^2\Omega_{m}\,\Theta\,\tilde{\delta}\,\ell =0\,,
\end{equation}
or 
\begin{equation}
\frac{d^2\ell}{d \tau^2}+ \Theta\,\tilde{\delta}\,\ell =0\,,
\end{equation}
where 
\begin{equation}
d\tau=H_0\sqrt{2\Omega_{m}}d\eta
=-\frac{\sqrt{2\Omega_{m}}\mathbf{d}z}{\sqrt{\Omega_\Lambda+\left[\Omega_k+\Omega_m(1+z)\right](1+z)^2}}
=-\frac{\sqrt{2}\mathbf{d}z}{\sqrt{\Omega_m^{-1}-1+(1+z)^3}}.
\end{equation}
For periodical line-of-sight distribution of voids we adopt \cite{GK2} 
\begin{equation}\label{period1}
\tilde{\delta}(\tau+\tau_k+\tau_\omega)=\tilde{\delta}(\tau) =
    \begin{cases}
      -\kappa^2 & 0< \tau < \tau_\kappa\\
      \ \omega^2 & \tau_\kappa < \tau < \tau_\kappa+\tau_\omega\, ,
    \end{cases}    
\end{equation}
and
$$
\Theta=1+\nu(\tau)\,,
$$
where $\nu$ is a stationary process with $\langle\nu(\tau)\rangle=0$ and auto-correlation function 
$$\Gamma(\tau)=\langle\nu(\tilde\tau)\nu(\tilde\tau+\tau)\rangle=\sigma^2\delta(\tau)$$. 

Then we arrive at the matrix equation

\begin{equation}\label{average2}
    \frac{d}{d\tau}\begin{pmatrix}
\langle \ell^2\rangle\\
\langle\dot{\ell}^2\rangle\\
\langle\ell\dot\ell\rangle
\end{pmatrix}
=
\begin{pmatrix}
0 & 0 & 2\\
\sigma^2\tilde\delta^2 & 0 & -2\tilde\delta\\
-\tilde\delta & 1 & 0
\end{pmatrix}
\begin{pmatrix}
\langle\ell^2\rangle\\
\langle\dot{\ell}^2\rangle\\
\langle\ell\dot\ell\rangle
\end{pmatrix}\,.
\end{equation}

The matrix of transformation $f$ after period $\tau_\kappa+\tau_\omega$  is 
\begin{equation}
f(\alpha)=e^{B\tau_\omega}e^{A\tau_\kappa}\left(\text{\it I}+ \alpha
\left[\kappa^4\int_0^{\tau_\kappa}e^{-As}Je^{As}ds
       +\omega^4 e^{-A\tau_\kappa}\left(\int_0^{\tau_\omega}e^{-Bs}Je^{Bs}ds\right)e^{A\tau_\kappa}\right]\right) + o(\alpha),
\end{equation}
where $\alpha=\sigma^2$ and 
$$
A=
\begin{pmatrix}
0 & 0 & 2\\
0 & 0 & 2\kappa^2\\
\kappa^2 & 1 & 0
\end{pmatrix}\ , \qquad
B=
\begin{pmatrix}
0 & 0 & 2\\
0 & 0 & -2\omega^2\\
-\omega^2 & 1 & 0
\end{pmatrix}\ , \qquad
J=
\begin{pmatrix}
0 & 0 & 0\\
1 & 0 & 0\\
0 & 0 & 0
\end{pmatrix}\ .
$$
Denoting as $\lambda_1$, $\lambda_2$, and $\lambda_3$ the eigenvalues of $f(\alpha)$, we will have for the distortion of the flow after $n$ line-of-sight periods
\begin{equation}
    \beta(n)=\left[\frac{\text{min}\{|\lambda_1|,|\lambda_2|,|\lambda_3|\}}    {\text{max}\{|\lambda_1|,|\lambda_2|,|\lambda_3|\}}\right]^{n/2}\,.
\end{equation}

For $f(0)$ we will have $\lambda_1=1$, $\lambda_2=\mu$, and $\lambda_3=\mu^{-1}$, where
\begin{itemize}
    \item $\mu$ is real and $0<\mu\le1$, or
    \item $\mu$ is complex and $|\mu|=1$,
\end{itemize}
and 
\begin{equation}
    \mu=\tfrac{1}{4}\left(b -\sqrt{b^2-4}\right)^2,
\end{equation}
where
\begin{equation}
    b=2 \cosh (\kappa\tau_\kappa) \cos (\omega \tau_\omega) 
    +\left(\frac{\kappa}{\omega } -\frac{\omega}{\kappa }\right)  \sinh (\kappa \tau_\kappa) 
    \sin (\omega \tau_\omega)\,,
\end{equation}
and
\begin{equation}
\text{tr}(f(0))=b^2-1\,.
\end{equation}

In addition, it can be shown that, if $\sigma^2\left|\text{tr}\left(f'(0)\right)\right|\ll|b^2-4|$, then
\begin{equation}
    \beta(n)\approx\begin{cases}
    1-\frac{3}{2}n\sigma^2 
    \left|\frac{\text{tr}\left(f'(0)\right)}{{b^2-4}}\right|\,, & \text{if}\qquad |b|<2\\
    \mu^n &  \text{if}\qquad |b|>2
    \end{cases}\,.
\end{equation}

Then for given $z$, $n$ is given by
\begin{equation}
n(z)=\frac{\sqrt{2}}{\tau_\kappa+\tau_\omega}\int_0^z\left[\Omega_m^{-1}-1+(1+\xi)^3\right]^{-1/2}d\xi\ .
\end{equation}

\section{Distortion's sensitivity on void/wall scales}

For $\kappa\tau_\kappa\ll1$, $\omega\tau_\omega\ll1$ one finds the following 

\begin{equation}
    \mu \approx 1-2 \sqrt{(\tau_\kappa +\tau_\omega) \left(\kappa ^2 \tau_\kappa -\omega ^2\tau_\omega\right)}\,.
\end{equation}

\begin{equation}
\text{tr}(f'(0))=
\kappa^4\text{tr}\left(e^{B\tau_\omega}e^{A\tau_\kappa}\int_0^{\tau_\kappa}e^{-As}Je^{As}ds\right)
+\omega^4 \text{tr}\left(e^{A\tau_\kappa}e^{B\tau_\omega}\int_0^{\tau_\omega}e^{-Bs}Je^{Bs}ds\right)\,,
\end{equation}

\begin{equation}
\text{tr}(f'(0))\approx
\tau_\kappa\kappa^4\text{tr}\left(e^{B\tau_\omega}Je^{A\tau_\kappa}\right)
+\tau_\omega\omega^4 \text{tr}\left(e^{A\tau_\kappa}Je^{B\tau_\omega}\right)
\approx (\tau_\kappa+\tau_\omega)^2(\kappa^4\tau_\kappa+\omega^4\tau_\omega)\,,
\end{equation}

\begin{equation}
    4 - b^2 \approx 4 (\tau_\kappa +\tau_\omega) \left(-\kappa ^2 \tau_\kappa + \omega ^2 \tau_\omega\right)>0\,.
\end{equation}
Hence
\begin{equation}
\frac{\text{tr}(f'(0))}{4 -b^2}
\approx \frac{\kappa^4\tau_\kappa+\omega^4\tau_\omega}
{4  \left(-\kappa ^2 \tau_\kappa + \omega ^2 \tau_\omega\right)}(\tau_\kappa+\tau_\omega)\,.
\end{equation}

We can rewrite the equations using the following physical terms
\begin{eqnarray}
&T=\tau_\kappa+\tau_\omega\\
&\langle\delta_0\rangle
=\frac{1}{T}\int_0^T\delta_0 d\tau 
= \frac{1}{T}\left(-\kappa^2\tau_\kappa +\omega^2\tau_\omega\right)\\
&\langle\delta_0^2\rangle
=\frac{1}{T}\int_0^T\delta_0^2 d\tau 
= \frac{1}{T}\left(\kappa^4\tau_\kappa +\omega^4\tau_\omega\right)
\end{eqnarray}

\begin{eqnarray}
\beta(n)\approx
\begin{cases}
1-2n \sqrt{-T\langle\delta_0\rangle}\ & \langle\delta_0\rangle<0\\
\\
1-\tfrac{3}{8}n\sigma^2\,T
\langle\delta_0^2\rangle/\langle\delta_0\rangle\ & \langle\delta_0\rangle>0
\end{cases}
\end{eqnarray}
Where $n$ is given by
\begin{equation}
n(z)=\frac{\sqrt{2}}{T}\int_0^z\left[\Omega_m^{-1}-1+(1+\xi)^3\right]^{-1/2}d\xi\ .
\end{equation}

From here, we obtain the distortion $\beta$'s dependency on the redshift, Fig. 1, and mean under-density parameter $\langle\delta_0\rangle$, Fig. 2ab. It is seen that the distortion is sensitive to void/wall scale ratio, e.g. at 0.1 of that ratio it does show variation on the redshift, while at higher ratio it remains nearly unchanged. 

These clearly illustrates the main outcome of the above analysis, namely, the sensitivity of the distortion caused by the hyperbolicity of voids on the observables characteristics of the voids and walls.


\begin{figure}[h]
\caption{Distortion $\beta$ vs the redshift $z$ for different values of
the ratio void/wall scales.}
\centering
\includegraphics[]{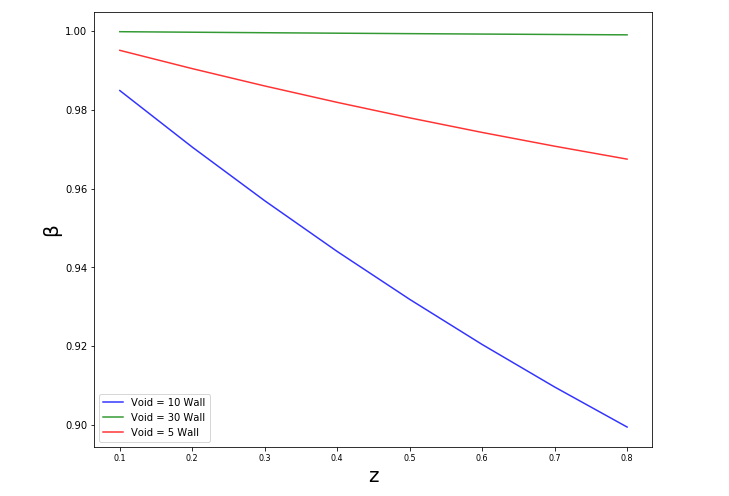}
\end{figure}

\begin{figure*}[h!]
\caption{Distortion $\beta$ vs the positive \textbf{(a)} and negative \textbf{(b)} values of the mean density parameter $\langle\delta_0\rangle$ for redshift $z=0.8$.}
\centering
\begin{subfigure}{0.5\textwidth}
\centering
\includegraphics[width=8.4cm]{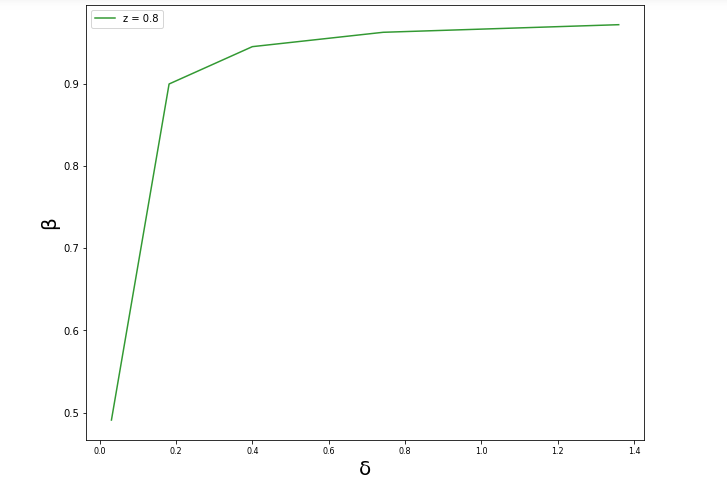}
\caption{}
\end{subfigure}%
\begin{subfigure}{0.5\textwidth}
\centering
\includegraphics[width=9cm]{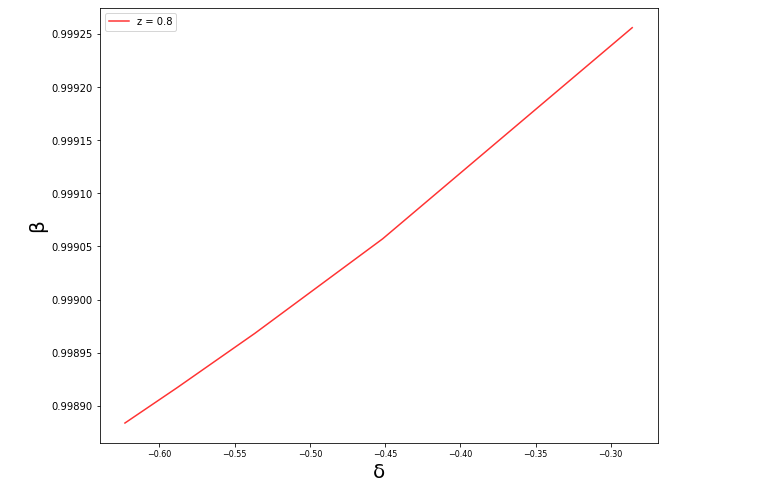}
\caption{}
\end{subfigure}
\end{figure*}


\section{Conclusion}

The hyperbolicity of photon beams i.e. their deviation is induced by cosmic voids as of low density regions in the large-scale Universe. The hyperbolicity will introduce distortion in the statistical features of galactic surveys, namely, distorting the two-point correlation functions \cite{SK}. The distortions of correlation function due to coordinate shifts in tangential directions have been detected, e.g. \cite{Pea,Guz}.  That distortion has to reflect e.g. the contributions of the peculiar motions of galaxies, including the Kayser-distortion due to the infall of galaxies toward the galaxy cluster center \cite{Mc,RG}. 
For example, the survey of 10,000 galaxies up to redshift $z=0.8$ had revealed a distortion $\beta \simeq 0.7$  \cite{Guz}.  That distortion of the tangential shift of galactic coordinates, if mainly due to the void hyperbolicity, would imply e.g. $N=6$ line-of-sight voids of mean diameter $D=50Mpc$ and mean density parameters of the walls of 4 Mpc mean size and voids, $\tilde\delta_{\text{Wall}}=10\,, \tilde\delta_{\text{Void}}=-0.8$, respectively \cite{SK}.

Another possible observational indications for the hyperbolicity are the Cosmic Microwave Background temperature anisotropy map distortions. Among the latter is the non-Gaussian anomaly of the Cold Spot. The analysis of its microwave pixelized maps revealed the predicted features of hyperbolicity of voids which hence supports the void nature of the Cold Spot \cite{spot}, and that conclusion was confirmed by subsequent 3D galactic survey \cite{Sz}.

We now revealed a property which can have principal role for galaxy survey data analysis, i.e. the sensitivity of the hyperbolicity signature on the observable parameters, the cumulative number of the line-of-sight voids and, in particular, on the ratio of the void/wall scales.  This opens a direct way to reveal the number distribution, and in particular, the physical scales of the cosmic voids as of key structures in the large-scale matter distribution in the Universe.

\section{Acknowledgments}

M.S. acknowledges the support of Armenian SCS grant 20RF-142.

\end{document}